\def\be{\begin{equation}}
\def\ee{\end{equation}}
\def\ba{\begin{eqnarray}}
\def\ea{\end{eqnarray}}
\def\l{\label}
\def\n{\nonumber \\}
\def\b{\bibitem}

\documentstyle[12pt]{article}
 \begin{document} 
\title{Generalized factorial moments}

\author{A.Bialas \\ M.Smoluchowski Institute of Physics \\Jagellonian
University, Cracow\thanks{Address: Reymonta 4, 30-059 Krakow, Poland;
e-mail:bialas@th.if.uj.edu.pl;}}
\maketitle

\begin{abstract} It is shown that the method of eliminating the
statistical fluctuations from event-by-event analysis proposed
recently by Fu and Liu can be rewritten in a compact form involving the
generalized factorial moments.

\end{abstract}
 
{\bf 1.} The  factorial moments of the multiplicity distribution
\ba
F_k\equiv <n(n-1)...(n-k+1)>    \l{1}
\ea
where $<...>$ denotes an average over events, 
proved to be a very useful tool in investigations of the multiplicity
fluctuations. The main reason of this success was the observation
\cite{bp} that the measurement of $F_k$ gives a direct access to the
"dynamical" fluctuations of the multiplicity.

The problem of separation of the "dynamical" and "statistical"
multiplicity fluctuations can be formulated as follows \cite{bp}.
 Assume that the
average multiplicity in a certain phase space region undergoes
"dynamical" fluctuations with the probability distribution $W(\bar{n})
d\bar{n}$ where $\bar{n}$ is the average multiplicity in this region. At
fixed $\bar{n}$ there are additional "statistical" fluctuations because
the actually observed number of particles obviously cannot be identical
with $\bar{n}$. Assuming that these statisticcal fluctuations do not
introduce new correlations in the system, we conclude that they
must be in the form of the Poisson distribution. Consequently, the
actually observed multiplicies are distributed according to
\ba
P(n) = \int d\bar{n} W(\bar{n}) e^{-\bar{n}} \frac{\bar{n}^n}{n!}     \l{2}
\ea
A simple calculation shows that the {\it factorial} moments of this
 distribution $P(n)$ are equal to  the {\it normal} moments of $W(\bar{n})$:
\ba
F_k=\sum_nn(n-1)...(n-k+1)P(n) = \int d\bar{n} \bar{n}^k W(\bar{n})  \l{3}
\ea
and thus indeed, $F_k$ gives a direct access to the {\it dynamical}
 distribution $W(\bar{n})$.

At this point it is also useful to recall that the factorial moments are
simply related to the integrals of the inclusive particle densities:
\ba
<n(n-1)...(n-k+1)> =\int dp_1dp_2...dp_k\rho(p_1,p_2,...,p_k) \l{4}
\ea
This formula  relates  multiplicity fluctuations
(as expressed by the factorial moment on the L.H.S.) and multiparticle
correlations (as expressed by the integral on the R.H.S.). Note also that
in absence of correlations between particles we obtain $F_k=\bar{n}^k$.
 
{\bf 2.} Recently, Fu and Liu \cite{fl} proposed an extension of this method
which can be used to eliminate
statistical fluctuations from distributions of other dynamical
quantities. In the present note we show that their result can be
elegantly formulated in terms of the {\it generalized factorial
moments}.

Following \cite{bk}, we are going to study 
fluctuations (in a given phase-space region) of 
 an {\it extensive}  quantity $X$ defined for each event as
\ba
X= \sum_{i=1}^n x_i  \l{5}
\ea
where $i=1,...n$ labels the particles in the event and $x_i$ is a
dynamical variable which may depend on 
momentum and other quantum numbers of  particle $i$. 
Some examples   of $x_i$
are charge,  momentum, or energy of the particle. Note that taking 
$x_i\equiv 1$
we obtain $X=n$, i.e., the problem is reduced to the previous one.

Let us now  define the  generalized factorial moments as

\ba
F_k[X] \equiv <[X-(k-1)\hat{x}]...[X-\hat{x}]X> \l{7}
\ea
where $\hat{x}$ is the operator acting on the variable $X$ in the
following way:
\ba
[\hat{x}]^lX  = X_{l+1}   \l{8}
\ea                                                                                 
with
\ba
X_l \equiv \sum_{i=1}^n [x_i]^l,\;\;\;\;\;X_1 \equiv X .  \l{9}
\ea

To see the physical interpretation of $F_k[X]$, let us
assume -in the spirit of the argument which led to the Eq. (\ref{2})-
that the "dynamical" fluctuations of the average values  $\bar{n}$ and 
$\bar{x}$ in a given
phase-space region are described by the distribution
 $W(\bar{n},\bar{x})d\bar{n}d\bar{x}$. To
obtain the actual distribution of $n$ and $x$, it is necessary to add
"statistical" fluctuations. Demanding again that they do not introduce
additional correlations in the system, we arrive at the formula
\ba
P(n,x_1,...,x_n)dx_1...dx_n= \int d\bar{n}d\bar{x}W(\bar{n},\bar{x})
e^{-\bar{n}}\frac{\bar{n}^n}{n!} 
p(x_1,\bar{x})dx_1...p(x_n,\bar{x})dx_n     \l{10}
\ea
where $p(x,\bar{x})dx$ is the distribution of $x$ at a fixed $\bar{x}$.

Using this we obtain
\ba
<X>=\int d\bar{n}d\bar{x}W(\bar{n},\bar{x})\bar{n}\bar{x}; \n
<X^2>=\int d\bar{n}d\bar{x}W(\bar{n},\bar{x})
 [(\bar{n}\bar{x})^2 \!+\! \bar{n}\bar{x^2}]=
\int d\bar{n}d\bar{x}  W(\bar{n},\bar{x})(\bar{n}\bar{x})^2 \!+\!<X_2>;\n
<X^3>=\int d\bar{n}d\bar{x}W(\bar{n},\bar{x}) [(\bar{n}\bar{x})^3 \!+\!
3\bar{n}\bar{x}\bar{n}\bar{x^2} \!+\!\bar{n}\bar{x^3}]=\n=
 \int d\bar{n}d\bar{x}W(\bar{n},\bar{x})(\bar{n}\bar{x})^3
\!+\!3[<XX_2>\!-\!<X_3>] \!+\!<X_3>;\n
<X^4>= \int d\bar{n}d\bar{x} W(\bar{n},\bar{x})[(\bar{n}\bar{x})^4\!+\!
6(\bar{n}\bar{x})^2\bar{n}\bar{x^2}\!+\!3(\bar{n}\bar{x^2})^2\!+\!4
\bar{n}\bar{x}\bar{n}\bar{x^3} \!+\!\bar{n}\bar{x^4}]=\n=
\int d\bar{n}d\bar{x} W(\bar{n},\bar{x})(\bar{n}\bar{x})^4+\n
6[<X^2X_2>\!-\!2(<XX_3>\!-\!<X_4>)\!-\!(<(X_2)^2>\!-\!<X_4>)
\!-\!<X_4>]\n+
3(<(X_2)^2>\!-\!<X_4>) \!+\!4(<XX_3>\!-\!<X_4>)\!+\!<X_4>   \l{11}
\ea
and similarly for higher moments. Here $\bar{x^k}= \int dx x^kp(x,\bar{x})$.

 Comparing (\ref{11}) with (\ref{7}) we conclude  that
\ba
F_k[X] = \int d\bar{n}d\bar{x}W(\bar{n},\bar{x})(\bar{n}\bar{x})^k   \l{12}
\ea
which shows that, indeed, the generalized factorial moment $F_k[X]$
measures directly the k-th moment of the {\it dynamical}
 distribution 
\ba
W(\bar{X})\equiv \int d\bar{n}d\bar{x}W(\bar{n},\bar{x})\delta(\bar{X}-
\bar{n}\bar{x})   \l{13}
\ea

{\bf 3.} It was shown in
\cite{bk} that the moments of $X$ can be expressed in terms of the
 finite number of the integrals of the products of {\it inclusive}
particle densities and powers of $x$, e.g.,
\ba
<X> = \int dp \rho(p) x(p) ;    \n
<X^2> = \int dp_1 dp_2 \rho(p_1,p_2) x(p_1)x(p_2) \!+\! \int dp \rho(p)
[x(p)]^2 ;\n
<X^3>= \int dp_1 dp_2 dp_3 \rho(p_1,p_2,p_3) x(p_1)x(p_2)x(p_3) \!+\!\n\!+\!
3\int dp_1 dp_2 \rho(p_1,p_2) x(p_1)[x(p_2)]^2 \!+\!
\int dp \rho(p) [x(p)]^3
\l{6}
\ea
and similar but more complicated relations for higher moments. 

Observing that
\ba
\int dp_1 dp_2 \rho(p_1,p_2)
[x(p_1)]^{s_1}[x(p_2)]^{s_2}=<X_{s_1}X_{s_2}>\!-\!<X_{s_1\!+\!s_2}>   \l{15}
\ea 
one sees that (\ref{6}) can be summarized in the single formula
\ba
F_k[X] = \int dp_1...dp_k \rho(p_1,...,p_k)x(p_1)...x(p_k) \l{14}
\ea
which is a generalization of (\ref{4}) and relates the fluctuations of
the variable $X$ with the correlations in the system. Although we have only
shown that (\ref{14}) is valid for $k=1,2,3$, it is not
difficult to see that it is actually valid for any $k$.

{\bf 3.} In summary, it was shown that the recently proposed method of
eliminating the statistical fluctuations from event-by-event
analysis \cite{fl} can be rewritten in a compact form in terms of
the generalized factorial moments. It was also shown that the well-known
relation between factorial moments and integrals of inclusive
multiparticle densities can be extended  to the generalized
factorial moments.

\vspace{0.3cm}

 {\bf Acknowledgements}

\vspace{0.3cm} I would like to thank J.Fu and L.Liu for sending me their
unpublished result which triggered this investigation. It was supported
in part by the by Subsydium of Foundation for Polish Science NP 1/99 and
by the Polish State Commitee for Scientific Research (KBN) Grant No 2
P03 B 09322.

\end{document}